\documentclass[aps,prl,reprint,superscriptaddress]{revtex4-2}

\usepackage{graphicx}
\usepackage{hyperref}
\usepackage{dcolumn}
\usepackage{bm}
\usepackage{color}
\usepackage{braket}
\usepackage{blindtext}
\usepackage{ragged2e} 
\usepackage{lipsum}
\usepackage{comment}

\DeclareUnicodeCharacter{2212}{-}

\begin{document}

\preprint{APS/123-QED}

\title{Constriction-induced modulation of charging energy in a quantum Hall cavity}

\author{Emily Hajigeorgiou}
\affiliation{Institute of Physics, École Polytechnique Fédérale de Lausanne (EPFL), CH-1015 Lausanne, Switzerland}
\author{Arup Kumar Paul}
\affiliation{Braun Center for Submicron Research, Department of Condensed Matter Physics, Weizmann Institute of Science, Rehovot 7610001, Israel}
\author{Mario Di Luca}
\affiliation{Institute of Physics, École Polytechnique Fédérale de Lausanne (EPFL), CH-1015 Lausanne, Switzerland}
\author{Vladimir Umansky}
\affiliation{Braun Center for Submicron Research, Department of Condensed Matter Physics, Weizmann Institute of Science, Rehovot 7610001, Israel}
\author{Moty Heiblum}
\affiliation{Braun Center for Submicron Research, Department of Condensed Matter Physics, Weizmann Institute of Science, Rehovot 7610001, Israel}
\author{Mitali Banerjee}
\email{mitali.banerjee@epfl.ch}
\affiliation{Institute of Physics, École Polytechnique Fédérale de Lausanne (EPFL), CH-1015 Lausanne, Switzerland}
\affiliation{Center for Quantum Science and Engineering (QSE Center), École Polytechnique Fédérale de Lausanne (EPFL), CH-1015 Lausanne, Switzerland}

\begin{abstract}
Electronic Fabry–Pérot interferometers (FPIs) operating in the fractional quantum Hall regime are a key platform for probing anyonic braiding statistics, yet interpreting their interference signals is complicated by Coulomb charging effects, which are commonly treated as parasitic, static properties governed by the cavity's geometry and electrostatics. Here, using a gate-defined quantum Hall cavity tuned to the Coulomb-dominated regime, we demonstrate that the charging energy is in fact strongly and non-monotonically modulated by the magnetic field, varying by up to 60\% over a range of only 100~mT. The effect appears exclusively when the quantum point contacts (QPCs) forming the cavity are weakly pinched off, i.e., in the strong cavity-to-lead coupling regime. By correlating the charging energy modulation with the QPC magneto-conductance, we attribute this behavior to field-dependent changes in local compressibility and electrostatic screening between the cavity and the leads, driven by the formation of incompressible fractional quantum Hall states within the constrictions. This result establishes QPC constrictions of quantum Hall cavities as active electrostatic elements rather than passive boundaries, revealing a dynamic screening mechanism, with direct consequences for the interpretation of interference measurements and the extraction of anyonic statistics.
\end{abstract}

\maketitle

\section*{Introduction}

Electronic cavities in the quantum Hall regime provide a platform that can be tuned continuously~\cite{Kim2024RonenNanotech,RoosliThesis,Halperin2011PRB,Stern2010PRB} between two limits: a Fabry-Pérot interferometer (FPI), governed by phase-coherent edge-state interference, and a quantum dot, dominated by charge quantization. The crossover between these regimes is controlled by several parameters, most notably the Coulomb interactions within the cavity. In the weak interaction limit, the system approaches the Aharonov-Bohm (AB) regime, characterized by a fixed interference area and continuous phase evolution with magnetic field \cite{Halperin1982QHE}. As Coulomb interactions increase, they induce partial area compensation (“area breathing”)~\cite{Sivan2016NatCom}, leading to an inversion of the constant-phase slope in the \(V_{PG} - B\) plane~\cite{Halperin2011PRB,Zhang2009_PRB} while retaining an interferometric character. In the extreme Coulomb-dominated (CD) limit, charge inside the cavity is strictly quantized, suppressing continuous magnetic field evolution and giving rise to characteristic vertical stripe patterns in the \(V_{PG} - B\) plane associated with discrete charging events~\cite{Ofek2010PNAS,Sivan2016NatCom}.

In Coulomb-dominated FPIs, interactions between the interfering edge and localized bulk charges (known as `bulk-edge' coupling) obscure the intrinsic interference phase, hindering the direct observation of anyonic statistics~\cite{Halperin2011PRB,Rosenow2007PRL,SimonRosenow2015PRL,Nakamura2022NatCom}. This challenge has motivated considerable effort to reduce these interactions in GaAs devices, for example, by increasing the cavity size~\cite{Zhang2009_PRB}, introducing electrostatic screening via top gates~\cite{McClure2009PRL,Zhang2009_PRB,Ofek2010PNAS}, adding a central ohmic contact~\cite{Choi2015NatCom,Sivan2018PRB,Sivan2016NatCom}, and engineering additional screening layers in the heterostructure~\cite{Nakamura2019NatPhys}. These approaches have been instrumental in accessing the AB regime, with the screening-well design enabling a small, coherent interferometer that yielded the first direct observation of anyonic braiding statistics~\cite{Nakamura2020Anyonic}. More recently, extending these concepts to graphene-based systems, devices incorporating screening gates have emerged as a promising platform, and have already demonstrated reliable observation of anyonic statistics~\cite{Kim2024RonenNanotech,Ronen2021Nanotech,Deprez2021Nanotech,Zhao2022,Fu2023,Werkmeister2024StronglyCoupled,Samuelson2025PRL,Kim2026Nature,Samuelson2026PRX,christina2026arxiv,Ronen2026arxiv}.

While reducing charging effects is crucial for achieving fully coherent AB-dominated FPIs, most approaches focus on mitigating bulk--edge coupling, which is regarded as the primary source of Coulomb-dominated behavior. This is motivated by the relation $E_C=e^2/C_\Sigma$, which suggests increasing the capacitance to suppress charging effects. However, $C_\Sigma$ is the total capacitance of the cavity, including coupling to the leads, and is often assumed to be independent of the magnetic field. The coupling to the leads is controlled by the quantum point contacts QPCs, which are primarily regarded as tunable beam splitters. Their role in determining the electrostatic environment of the cavity is often overlooked. If QPCs influence the cavity capacitance, their electrostatic contribution should be most apparent in the charging energy itself.

To isolate the charging physics from the additional complexities of interference, we study a quantum Hall cavity tuned into the extreme Coulomb-dominated regime~\cite{Ofek2010PNAS,Sivan2016NatCom}. In this limit, the charging energy can be measured directly and related to the total cavity capacitance, while the underlying geometry remains closely connected to that of an FPI. By tracking the charging energy as a function of magnetic field at a fixed cavity charge state, we find that $E_C$, and by extension the total capacitance $C_\Sigma$, can exhibit a pronounced magnetic-field dependence. Specifically, the charging energy displays a strong, non-monotonic evolution that emerges only as the QPCs become more transparent---precisely in the regime relevant for interference experiments---increasing by up to 60\% over a magnetic-field range of only 100~$\mathrm{mT}$. We attribute this behavior to magnetic-field-dependent capacitance changes associated with the formation of incompressible fractional quantum Hall states within the QPCs. Our results establish the electrostatic role of QPCs as a significant and previously overlooked contribution to the charging physics of quantum Hall interferometers, demonstrating that the surrounding electrostatic environment can evolve during magnetic-field sweeps with direct implications for the interpretation of FPI experiments.

\section*{Device and setup}

\begin{figure}[htb!]
 \includegraphics[width = 0.35\textwidth]{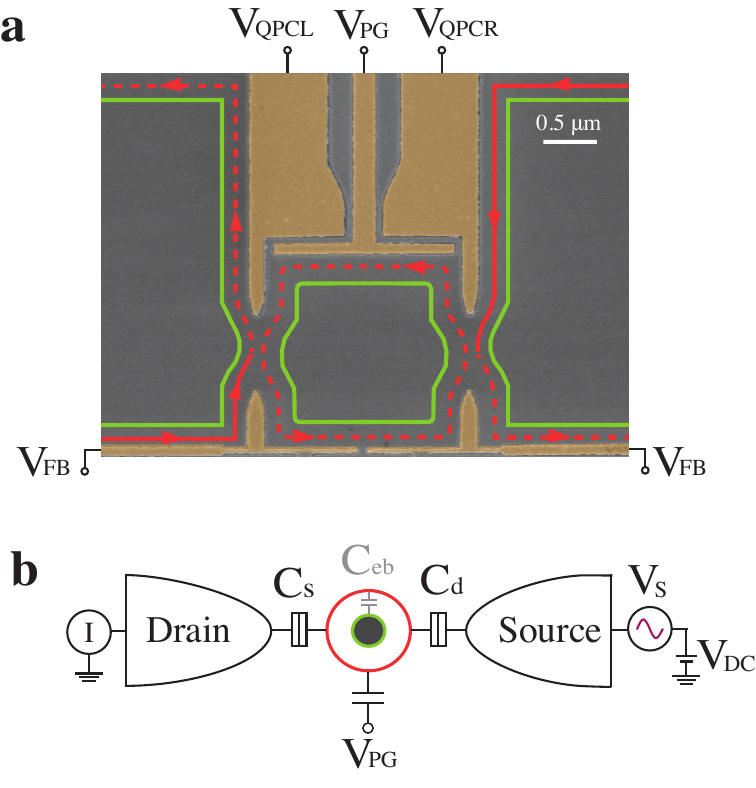}
 \begin{center}

 \caption{\textbf{Device picture and schematic.} 
\textbf{a}, False-colored SEM image of the \(2 \times 2\,\mu\mathrm{m}^2\) cavity. The fishbone gates are biased at \(V_{\mathrm{FB}} = -0.6\,\mathrm{V}\), and the left, right QPC, and plunger gate voltages are denoted \(V_{\mathrm{QPCL}}\), \(V_{\mathrm{QPCR}}\), and \(V_{\mathrm{PG}}\). The edge-state structure at \(\nu = 2\) is shown: the inner edge is fully reflected (light green), while the outer edge is partially transmitted at the QPCs (red). \textbf{b}, Schematic of the cavity and its electrostatic environment. A small AC excitation \(V_s\) is applied on top of a DC bias \(V_{\mathrm{DC}}\) at the source, and the resulting AC current is measured at the drain. The dot is tunnel-coupled to the source and drain via the QPCs, with capacitances \(C_{\mathrm{S}}\) and \(C_{\mathrm{D}}\), and capacitively coupled to the plunger gate (\(C_{\mathrm{PG}}\)). The bulk--edge capacitance $C_{\mathrm{eb}}$ (grey) is shown for illustration and does not enter the total capacitance \(C_\Sigma\) within the constant-interaction model.}
 \label{fig:1}
 \end{center}
\end{figure}

The device is fabricated on a GaAs/AlGaAs heterostructure hosting a two-dimensional electron gas (2DEG) with density \(\sim 1 \times 10^{11}~\mathrm{cm}^{-2}\). Metallic surface gates are negatively biased to deplete the 2DEG and define the cavity (lithographically defined to be \(2 \times 2~\mu\mathrm{m}^2\) as seen in Fig.~\ref{fig:1}a). Unless stated otherwise, measurements are performed at high perpendicular magnetic field in the integer quantum Hall regime at filling factor \(\nu = 2\), where two chiral edge states propagate along the sample boundaries~\cite{Halperin1982QHE,Chklovskii1992Edge}. The confinement potential also supports closed-edge trajectories within the cavity (Fig.~\ref{fig:1}a): the inner edge (green) is fully reflected at both QPCs, while the outer edge (red) is partially transmitted and contributes to transport. In all of our measurements, only the outer edge is partitioned. 

Two nominally identical devices were measured and exhibit qualitatively similar behavior.  The two samples were fabricated on adjacent chips of the same wafer and have identical cavity areas. Unless stated otherwise, all data shown in the main text correspond to device A (D-A), except for the magnetic-field dependence of the QPCs, which is presented for device B (D-B) in Fig.~\ref{fig:4}b,c. Following a density change observed in D-A, the full set of measurements was repeated on D-B (see Supplementary Information), yielding consistent results.

We perform two-terminal, voltage-biased measurements by applying a small AC excitation \(V_s\) at the source and measuring the resulting current at the drain (Fig.~\ref{fig:1}b), yielding the differential conductance \(dI/dV_s\), hereafter denoted as \(G\) in units of \(e^2/h\). The confinement potential is tuned to reach the extreme CD limit, where the cavity behaves as a large quantum dot. This is achieved by tuning the QPCs to strongly backscatter the outer edge. Scanning the left and right barrier voltages in Fig.~\ref{fig:3}a, results in conductance oscillations of the cavity near pinch-off. In the upper right corner, the oscillations are broadened by strong tunnel coupling to the leads, while making the barriers more negative (toward the lower left) leads to well-defined, predominantly temperature-broadened Coulomb blockade resonances.

\begin{figure*}[htb!]
 \includegraphics[width = 0.95\textwidth]{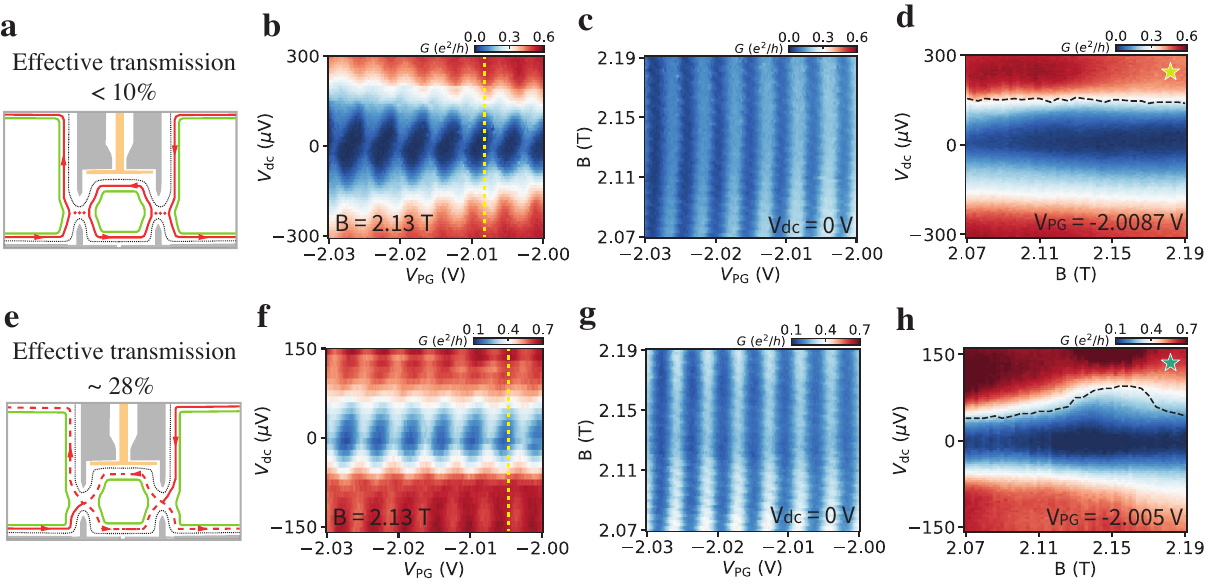}
 \begin{center}
 \caption{
 \textbf{a--d}, Effective transmission $<10\%$ (QPC setting as in Fig.~\ref{fig:3}e). 
\textbf{e--h}, Effective transmission $\approx 28\%$ (QPC setting as in Fig.~\ref{fig:3}f).  
\textbf{a,e}, Device schematics for the two regimes; only the outer edge is partitioned. The plunger gate (\(V_{\mathrm{PG}}\)) is highlighted in yellow, while all other gates (grey) are fixed. Black dotted lines indicate the confinement potential.  
\textbf{b,f}, Coulomb diamonds taken at 2.13 T; the yellow dashed line marks the fixed plunger-gate voltage used in \textbf{d,h}.  
\textbf{c,g}, Evolution of conductance oscillations with magnetic field taken at $V_{DC} = 0~V$.  
\textbf{d,h}, Conductance at fixed plunger-gate voltage (yellow dashed line in \textbf{b,f}), showing the evolution of a fixed charge state \(N\) with magnetic field and DC bias. The black dashed line traces the diamond height \(V_{\mathrm{DC}}(B)\), used to calculate the charging energy \(E_{\mathrm{c}}(B)\) ($=eV_{\mathrm{DC}}(B)$). The yellow (\textbf{d}) and cyan (\textbf{h}) stars correspond to the respective points in Fig.~\ref{fig:3}a. For details on how \(V_{\mathrm{DC}}(B)\) is extracted, see Supplementary. The DC bias is the fast axis in this measurement.
 }
 \label{fig:2}
 \end{center}
\end{figure*}

\section*{Results}

\subsection*{Charging behavior in weak and strong coupling regimes}

Figure~\ref{fig:2} compares transport through the cavity in two cavity--lead coupling regimes, distinguished by the transparency of the constrictions. Fig.~\ref{fig:2}a--d correspond to the weak-coupling regime, while Fig.~\ref{fig:2}e--h are obtained after symmetrically opening both barriers. The effective transmission, defined as the maximum zero-bias conductance of the Coulomb oscillations (Fig.~\ref{fig:2}a,e), serves as an empirical measure of the cavity--lead coupling since it increases with increasing transparency. 

We first consider the weak-coupling regime. Well-defined Coulomb diamonds emerge in Fig.~\ref{fig:2}b. The magnetic-field evolution of the Coulomb resonances at zero DC bias is shown in Fig.~\ref{fig:2}c. Over a range of \(\pm15\%\) around the center of the \(\nu=2\) plateau, the resonances exhibit no continuous magnetic-field evolution. No hysteresis is observed between upward and downward magnetic-field sweeps.

In the strongly CD regime (\(E_C \gg k_B T, \hbar \Gamma\)), the enclosed charge remains fixed between charging events~\cite{Ofek2010PNAS,Sivan2016NatCom}. Additionally, in the quantum Hall regime, varying \(B\) changes the Landau-level degeneracy and would therefore modify the enclosed charge at fixed area and filling factor \(\nu\) (\(N \sim \nu BA/\Phi_0\)). To avoid this energy cost, the cavity adjusts its effective area such that \(BA \approx \mathrm{constant}\). Consequently, as the magnetic field increases, the effective area continuously shrinks until one flux quantum is added, at which point an electron abruptly transfers from the bulk to the edge and the area re-expands. The resulting \(2\pi\) phase jump is not directly observable in transport. This phenomenon, commonly referred to as ``area breathing''~\cite{Sivan2016NatCom,Roosli2020PRB}, reflects electrostatic charge redistribution within the cavity rather than a rigid geometric motion of the edge.

\begin{figure*}[htb!]
 \includegraphics[width = 0.95\textwidth]{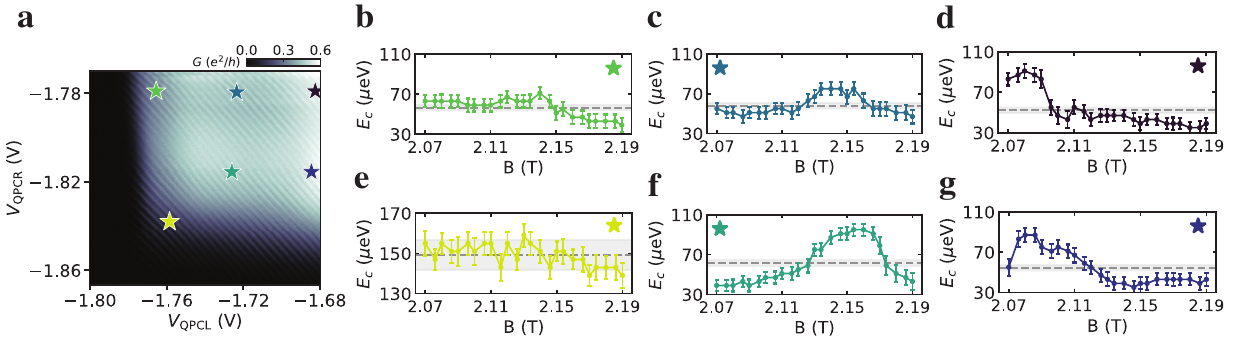}
 \begin{center}
 \caption{\textbf{a}, Conductance map as a function of left and right barrier voltages, with $V_{PG} = -2$ V. Each colored star corresponds to a trace in the following panels, labeled accordingly. \textbf{b--g}, Charging energy as a function of magnetic field, extracted from \(V_{\mathrm{DC}}(B)\) obtained from measurements similar to those shown in Fig.~\ref{fig:2}d,h. The grey dashed line indicates the mean, and the shaded region denotes the $\pm10\%$ interval.}
 \label{fig:3}
 \end{center}
\end{figure*}

We now turn to the strongly-coupled regime, with an effective transmission of \(28\%\). Compared to the weak-coupling regime, the charging energy is reduced, as evidenced by the smaller Coulomb diamonds in Fig.~\ref{fig:2}f relative to Fig.~\ref{fig:2}b. This reduction is expected since increasing the barrier transparency enhances the capacitive coupling between the cavity and the leads, thereby increasing the total capacitance and reducing \(E_C\).

Despite the stronger lead coupling, Coulomb-blockade diamonds with suppressed conductance inside the blockade region remain clearly visible. Moreover, the magnetic-field evolution of the zero-bias oscillations (Fig.~\ref{fig:2}g) remains qualitatively unchanged, indicating that the cavity is still strongly Coulomb dominated. In both regimes, an additional periodic modulation along the Coulomb peaks is observed as a function of magnetic field. Similar behavior has been reported previously~\cite{Roosli2020PRB,Roosli2021SciAdv} and attributed to internal charge rearrangements between compressible regions that conserve the total cavity charge.

\begin{figure}[htb!]
 \includegraphics[width = 0.48\textwidth]{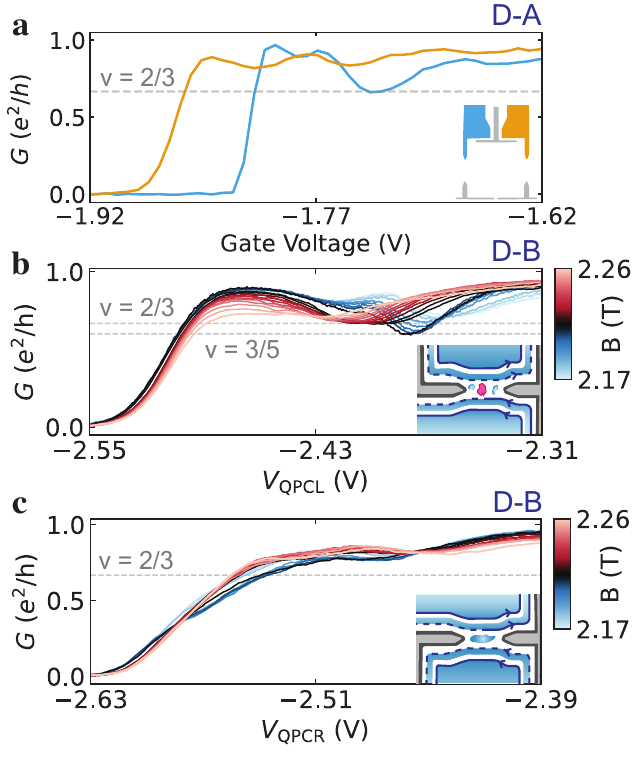}
 \begin{center}
 \caption{\textbf{a}, Conductance through the left (blue) and right (orange) QPCs of device A (D-A), measured while the opposite constriction fully transmits the outer edge mode. The left QPC exhibits a clear \(\nu = 2/3\) plateau. Data are taken at \(B = 2.13~\mathrm{T}\). 
\textbf{b}, Conductance through the left QPC of device B (D-B) as a function of magnetic field, with the right QPC fully open. A \(\nu = 3/5\) feature evolves into a pronounced \(\nu = 2/3\) plateau with increasing magnetic field. 
\textbf{c}, Conductance through the right QPC of device B (D-B) as a function of magnetic field, with the left QPC fully open. The inset schematics in \textbf{b,c} illustrate the QPC region in the two regimes: in \textbf{b}, an incompressible fractional quantum Hall state (pink) forms inside the constriction, while in \textbf{c} the QPC remains compressible.}
 \label{fig:4}
 \end{center}
\end{figure}

So far, the two coupling regimes appear qualitatively similar. To compare the two regimes quantitatively, we track the Coulomb-diamond apex (the point corresponding to the charging energy of the cavity) as a function of magnetic field. The plunger gate is fixed at the voltage indicated by the yellow dashed line in Fig.~\ref{fig:2}b, selecting a fixed charge state, and vertical DC-bias line cuts are measured as a function of magnetic field (Fig.~\ref{fig:4}d,h). This allows us to extract the diamond apex as a function of the magnetic field, \(V_{\mathrm{DC}}(B)\) (see Supplementary). In the weak-coupling regime, the extracted \(V_{\mathrm{DC}}(B)\) (black dashed line in Fig.~\ref{fig:2}d) remains approximately constant as a function of B.

In contrast, performing the same measurement in the higher-transmission regime yields the result shown in Fig.~\ref{fig:2}h, where \(V_{\mathrm{DC}}(B)\) exhibits a pronounced and non-monotonic magnetic-field dependence.

\subsection*{Magnetic field dependence of charging energy}

After extracting \(V_{\mathrm{DC}}(B)\), we determine the charging energy as \(E_C(B)=eV_{\mathrm{DC}}(B)\) (positive-bias branch, see Supplementary Information). The resulting \(E_C(B)\) is shown in Fig.~\ref{fig:3}b--g for the corresponding QPC combinations indicated by the colored stars in Fig.~\ref{fig:3}a. For the data discussed in Fig.~\ref{fig:2}, the extracted charging energy for effective transmission \(<10\%\) is shown in Fig.~\ref{fig:3}e, while the $28\%$ transmission case is shown in Fig.~\ref{fig:3}f. In the weak-coupling regime, the charging energy exhibits no systematic magnetic-field dependence within \(\pm10\%\) of its mean value and is therefore effectively constant over this range. In contrast, for more transparent barriers, \(E_C(B)\) varies non-monotonically by up to \(60\%\) over a magnetic-field range of only \(\sim100~\mathrm{mT}\). To verify that we are continuously probing the apex of the same Coulomb diamond as a function of B, we repeat Coulomb diamond measurements at several distinct magnetic-field values (see Supplementary Information).

The measurements above establish that the magnetic-field dependence of the charging energy depends strongly on barrier transparency. To isolate the role of each constriction, we repeat the measurement for all QPC combinations indicated in Fig.~\ref{fig:3}a. We exclude Fig.~\ref{fig:3}e from the following discussion, since it displays the expected nearly constant behavior.

We first analyze the effect of the right barrier. Moving vertically in Fig.~\ref{fig:3}a predominantly changes the transparency of the right QPC while keeping the left barrier approximately fixed. Comparing Fig.~\ref{fig:3}c and \ref{fig:3}f, the charging energy exhibits a maximum at nearly the same magnetic field in both cases, followed by a decrease away from this point. A similar correspondence is observed between Fig.~\ref{fig:3}d and \ref{fig:3}g, where \(E_C(B)\) peaks at a lower magnetic field and then saturates. Thus, varying the right barrier over a wide voltage range (\(\sim40~\mathrm{mV}\)) does not qualitatively alter the magnetic-field dependence of \(E_C(B)\).

In contrast, varying the left barrier leads to noticeably different behavior. Moving horizontally in Fig.~\ref{fig:3}a changes the transparency of the left QPC, and comparing Fig.~\ref{fig:3}b--d reveals qualitatively distinct magnetic-field evolution. The same behavior is seen between Fig.~\ref{fig:3}f and \ref{fig:3}g. These observations demonstrate that the nontrivial behavior of \(E_C(B)\) depends sensitively on the left constriction.

Taken together, these results indicate that the electrostatic environment governing the charging energy is controlled in a spatially dependent manner within the constrictions; predominantly the left one. Figure~\ref{fig:4}a shows the conductance through the two QPCs in device A, measured while the opposite constriction fully transmits the outer edge mode. The left QPC exhibits a pronounced \(\nu=2/3\) plateau, whereas the right QPC remains compressible. The appearance of hole-conjugate fractional quantum Hall (FQH) states in constrictions is widely attributed to edge reconstruction induced by soft confinement potentials~\cite{Kane1995Edgerec,Bid2010Edgerec,Sabo201Edgerec,wang2013PRLEdgerec}. In GaAs devices, reconstructed fractional edges are commonly associated with upstream neutral modes~\cite{Rajesh2019PRLNeutral,Inoue2014NatComNeutral,Inoue2013FracIQHE,GoldsteinGefen2016PRL,Rosenblatt2017}. Such effects are particularly pronounced in narrow constrictions, where sharper confinement enhances edge reconstruction and inter-edge coupling. The resulting abrupt spatial variation of the local filling factor increases the sensitivity of the conductance to small changes in gate voltage, thereby stabilizing and revealing weaker fractional quantum Hall states in transport~\cite{Baer2014QPC}.

In both devices A (Fig.~\ref{fig:4}a) and B (Fig.~\ref{fig:4}b), the left constriction exhibits a \(\nu=2/3\) plateau within the bulk \(\nu=2\) regime, while the right constriction remains compressible without observable fractional states. Figures~\ref{fig:4}b,c show the magnetic field evolution of the left and right constrictions in device B. Sweeping the magnetic field changes the local filling factor inside the constriction, allowing incompressible states to form when available, while localized states become populated as the Landau-level occupancy evolves. As seen in Fig.~\ref{fig:4}b, a \(\nu=3/5\) feature evolves into a pronounced \(\nu=2/3\) plateau over a narrow magnetic-field range in the left QPC.
 
This demonstrates that incompressible states can form within the constrictions while scanning the magnetic field in a small range, to characterize the cavity. This can modify the local screening between the cavity and the leads. Within the constant-interaction framework (see Supplementary), variations in \(E_C(B)\) directly reflect variations in the total capacitance,
\[
C_\Sigma(B)=C_\mathrm{s}(B)+C_\mathrm{d}(B)+C_\mathrm{PG}(B).
\]
A dominant magnetic-field contribution from \(C_\mathrm{PG}\) can be excluded since the gate periodicity remains unchanged over the investigated field range (\(C_\mathrm{PG}=e/\Delta V_\mathrm{PG}\), Figs.~\ref{fig:2}c,g). The remaining contributions therefore originate from the cavity--lead coupling, consistent with the strong dependence of \(E_C(B)\) on constriction transparency.

We therefore attribute the dominant magnetic-field dependence to variations in the effective cavity--lead capacitance, \(C_\mathrm{leads}(B)=C_\mathrm{s}(B)+C_\mathrm{d}(B)\), arising from changes in compressibility and screening within the constrictions. When an incompressible state forms inside a constriction, screening between the edge state arriving from the lead and the edge state inside the cavity is reduced, lowering the effective cavity--lead capacitance and increasing the charging energy. This is what we observe in the left QPC. In contrast, compressible constrictions provide more effective screening, increasing the total capacitance and reducing \(E_C\).

Importantly, the transition between incompressible states in the left constriction (Fig.~\ref{fig:4}b) occurs over the same magnetic-field range of 40~mT in which the charging energy exhibits pronounced maxima in Fig.~\ref{fig:3}c,d,f,g, directly correlating the evolution of \(E_C(B)\) with the formation of fractional states in the left constriction. 

Upon increasing the cryostat temperature to 200\,mK (see Supplementary), the $\nu=2/3$ plateau in the left QPC weakens but does not vanish. At the same time, the magnetic-field dependence of the charging energy is strongly suppressed compared to 10\,mK, tending toward its mean value. This result indicates that changes in QPC transmission, induced by thermal broadening, are correlated with modifications of the charging energy.

Finally, we note that related behavior has been reported in a graphene interferometer operated in the Aharonov--Bohm regime~\cite{Duprezthesis}. In that work, the constrictions were intentionally tuned to an incompressible fractional state, \(\nu=3/5\), while the bulk remained at \(\nu=2\). Under these conditions, the interference pattern in the \(V_{\mathrm{PG}}-B\) plane developed nearly vertical features, or equivalently, magnetic-field oscillations with an anomalously large period inconsistent with the interference area. In contrast, conventional AB behavior was recovered for other constriction settings. This sensitivity to the local state of the constrictions is consistent with our conclusion that constriction electrostatics can strongly modify the effective cavity environment.

\section*{Conclusions and Outlook}

We conclude that the charging energy of a quantum Hall cavity can exhibit a significant magnetic-field dependence arising from a field-dependent capacitance between the cavity and the leads. We attribute this behavior to changes in the compressibility within the QPCs, being tuned near an incompressible FQH state. While the magnitude of this effect is device dependent, set by the QPC design, potential softness, and local disorder, it is expected to become increasingly relevant as interferometers are scaled down to achieve better phase coherence, where narrower QPCs favor the formation of such incompressible regions.

More broadly, our results suggest that the charging energy of an interferometer should not necessarily be regarded as a fixed parameter during magnetic-field sweeps. Since phase jumps and interference patterns are commonly interpreted within models that assume a static electrostatic environment, a magnetic-field-dependent charging energy introduces an additional mechanism by which the interference phase can evolve. In particular, changes in the compressibility of the constrictions may modify the charging energy continuously or abruptly as fractional states form within the QPCs, potentially contributing to phase evolution that would otherwise be attributed solely to bulk quasiparticle dynamics. This finding motivates systematic characterization of QPC electrostatics as a function of magnetic field, and suggests that such measurements should become a standard part of the characterization of quantum Hall interferometers.

Part of the challenge in identifying this contribution is that it is not directly visible in standard interferometer characterization measurements. Instead, it becomes apparent only when the charging energy is explicitly traced as a function of the magnetic field. In the Aharonov--Bohm regime, the finite slope of the interference pattern in the \(V_{\mathrm{PG}}-B\) plane implies that sweeping the magnetic field at fixed plunger-gate voltage does not probe the same quantity. Rather, one must follow constant-phase trajectories (i.e., vary \(V_{\mathrm{PG}}\) and \(B\) simultaneously). In this regime, Coulomb diamonds are replaced by a checkerboard pattern arising from the interplay of interference and charging. A diagonal line cut along this pattern does not directly yield the charging energy; however, the geometry of the checkerboard, including the spacing and slope of its features, is influenced by \(E_C\) and thus encodes its magnetic-field dependence.

\begin{acknowledgments}
E.H. and M.D.L acknowledge funding from SNSF. M.B. acknowledges the support of the SNSF Eccellenza grant No. PCEGP2\_194528, and support from the QuantERA II Programme that has received funding from the European Union’s Horizon 2020 research and innovation program under Grant Agreement No 101017733. M.H. acknowledges the support of the Israel
Science Foundation, Grant No. 1510/22.
\end{acknowledgments}

\textbf{\begin{center}Data availability\end{center}}
The data supporting the findings of this study are available from the corresponding author upon reasonable request.

\bibliography{biblio}

@article{Werkmeister2024StronglyCoupled,
	author = {Werkmeister, Thomas and Ehrets, James R. and Ronen, Yuval and Wesson, Marie E. and Najafabadi, Danial and Wei, Zezhu and Watanabe, Kenji and Taniguchi, Takashi and Feldman, D. E. and Halperin, Bertrand I. and Yacoby, Amir and Kim, Philip},
	title = {{Strongly coupled edge states in a graphene quantum Hall interferometer}},
	journal = {Nat. Commun.},
	volume = {15},
	number = {6533},
	pages = {1--10},
	year = {2024},
	month = aug,
	issn = {2041-1723},
	publisher = {Nature Publishing Group},
	doi = {10.1038/s41467-024-50695-1}
}

@article{Ronen2021Nanotech,
	author = {Ronen, Yuval and Werkmeister, Thomas and Haie Najafabadi, Danial and Pierce, Andrew T. and Anderson, Laurel E. and Shin, Young Jae and Lee, Si Young and Lee, Young Hee and Johnson, Bobae and Watanabe, Kenji and Taniguchi, Takashi and Yacoby, Amir and Kim, Philip},
	title = {{Aharonov{\textendash}Bohm effect in graphene-based Fabry{\textendash}P{\ifmmode\acute{e}\else\'{e}\fi}rot quantum Hall interferometers}},
	journal = {Nat. Nanotechnol.},
	volume = {16},
	pages = {563--569},
	year = {2021},
	month = may,
	issn = {1748-3395},
	publisher = {Nature Publishing Group},
	doi = {10.1038/s41565-021-00861-z}
}

@article{Zhao2022,
  title = {Graphene-Based Quantum Hall Interferometer with Self-Aligned Side Gates},
  volume = {22},
  ISSN = {1530-6992},
  url = {http://dx.doi.org/10.1021/acs.nanolett.2c03805},
  DOI = {10.1021/acs.nanolett.2c03805},
  number = {23},
  journal = {Nano Letters},
  publisher = {American Chemical Society (ACS)},
  author = {Zhao,  Lingfei and Arnault,  Ethan G. and Larson,  Trevyn F. Q. and Iftikhar,  Zubair and Seredinski,  Andrew and Fleming,  Tate and Watanabe,  Kenji and Taniguchi,  Takashi and Amet,  Fran\c{c}ois and Finkelstein,  Gleb},
  year = {2022},
  month = Nov,
  pages = {9645–9651}
}

@article{Fu2023,
  title = {Aharonov–Bohm Oscillations in Bilayer Graphene Quantum Hall Edge State Fabry–Pérot Interferometers},
  volume = {23},
  ISSN = {1530-6992},
  url = {http://dx.doi.org/10.1021/acs.nanolett.2c05004},
  DOI = {10.1021/acs.nanolett.2c05004},
  number = {2},
  journal = {Nano Letters},
  publisher = {American Chemical Society (ACS)},
  author = {Fu,  Hailong and Huang,  Ke and Watanabe,  Kenji and Taniguchi,  Takashi and Kayyalha,  Morteza and Zhu,  Jun},
  year = {2023},
  month = Jan,
  pages = {718–725}
}

@article{Samuelson2025PRL,
  title = {Hard and Soft Phase Slips in a Fabry-P\'erot Quantum Hall Interferometer},
  author = {Samuelson, N. L. and Cohen, L. A. and Wang, W. and Blanch, S. and Taniguchi, T. and Watanabe, K. and Zaletel, M. P. and Young, A. F.},
  journal = {Phys. Rev. Lett.},
  volume = {134},
  issue = {25},
  pages = {256301},
  numpages = {6},
  year = {2025},
  month = {Jun},
  publisher = {American Physical Society},
  doi = {10.1103/kqct-3llf},
  url = {https://link.aps.org/doi/10.1103/kqct-3llf}
}

@article{Samuelson2026PRX,
  title = {Slow Quasiparticle Dynamics and Anyonic Statistics in a Fractional Quantum Hall Fabry-P\'erot Interferometer},
  author = {Samuelson, Noah L. and Cohen, Liam A. and Wang, Will and Blanch, Simon and Taniguchi, Takashi and Watanabe, Kenji and Zaletel, Michael P. and Young, Andrea F.},
  journal = {Phys. Rev. X},
  volume = {16},
  issue = {1},
  pages = {011062},
  numpages = {12},
  year = {2026},
  month = {Mar},
  publisher = {American Physical Society},
  doi = {10.1103/fwjg-mx9h},
  url = {https://link.aps.org/doi/10.1103/fwjg-mx9h}
}

@article{christina2026arxiv,
  doi = {10.48550/ARXIV.2603.11182},
  url = {https://arxiv.org/abs/2603.11182},
  author = {Henzinger,  Christina E. and Ehrets,  James R. and Fushio,  Rikuto and Dong,  Junkai and Werkmeister,  Thomas and Wesson,  Marie E. and Watanabe,  Kenji and Taniguchi,  Takashi and Vishwanath,  Ashvin and Halperin,  Bertrand I. and Yacoby,  Amir and Kim,  Philip},
  title = {Controlled localization of anyons in a graphene quantum Hall interferometer},
  journal = {arXiv preprint arXiv:2603.11182},
  publisher = {arXiv},
  year = {2026},
  copyright = {arXiv.org perpetual,  non-exclusive license}
}

@article{Ronen2026arxiv,
  doi = {10.48550/ARXIV.2603.11162},
  url = {https://arxiv.org/abs/2603.11162},
  author = {Kim,  Jehyun and Shaer,  Amit and Kumar,  Ravi and Ilin,  Alexey and Watanabe,  Kenji and Taniguchi,  Takashi and Stern,  Ady and Mross,  David F. and Ronen,  Yuval},
  title = {Selective braiding of different anyons in the even-denominator fractional quantum Hall effect},
  journal = {arXiv preprint arXiv:2603.11162},
  publisher = {arXiv},
  year = {2026},
  copyright = {arXiv.org perpetual,  non-exclusive license}
}

@article{Deprez2021Nanotech,
	author = {D{\ifmmode\acute{e}\else\'{e}\fi}prez, Corentin and Veyrat, Louis and Vignaud, Hadrien and Nayak, Goutham and Watanabe, Kenji and Taniguchi, Takashi and Gay, Fr{\ifmmode\acute{e}\else\'{e}\fi}d{\ifmmode\acute{e}\else\'{e}\fi}ric and Sellier, Hermann and Sac{\ifmmode\acute{e}\else\'{e}\fi}p{\ifmmode\acute{e}\else\'{e}\fi}, Benjamin},
	title = {{A tunable Fabry{\textendash}P{\ifmmode\acute{e}\else\'{e}\fi}rot quantum Hall interferometer in graphene}},
	journal = {Nat. Nanotechnol.},
	volume = {16},
	pages = {555--562},
	year = {2021},
	month = may,
	issn = {1748-3395},
	publisher = {Nature Publishing Group},
	doi = {10.1038/s41565-021-00847-x}
}

@article{Kim2024RonenNanotech,
	author = {Kim, Jehyun and Dev, Himanshu and Kumar, Ravi and Ilin, Alexey and Haug, Andr{\ifmmode\acute{e}\else\'{e}\fi} and Bhardwaj, Vishal and Hong, Changki and Watanabe, Kenji and Taniguchi, Takashi and Stern, Ady and Ronen, Yuval},
	title = {{Aharonov{\textendash}Bohm interference and statistical phase-jump evolution in fractional quantum Hall states in bilayer graphene}},
	journal = {Nat. Nanotechnol.},
	volume = {19},
	pages = {1619--1626},
	year = {2024},
	month = nov,
	issn = {1748-3395},
	publisher = {Nature Publishing Group},
	doi = {10.1038/s41565-024-01751-w}
}

@article{Kim2026Nature,
  title = {Aharonov–Bohm interference in even-denominator fractional quantum Hall states},
  volume = {649},
  ISSN = {1476-4687},
  url = {http://dx.doi.org/10.1038/s41586-025-09891-2},
  DOI = {10.1038/s41586-025-09891-2},
  number = {8096},
  journal = {Nature},
  publisher = {Springer Science and Business Media LLC},
  author = {Kim,  Jehyun and Dev,  Himanshu and Shaer,  Amit and Kumar,  Ravi and Ilin,  Alexey and Haug,  André and Iskoz,  Shelly and Watanabe,  Kenji and Taniguchi,  Takashi and Mross,  David F. and Stern,  Ady and Ronen,  Yuval},
  year = {2026},
  month = Jan,
  pages = {323–329}
}

@article{Nakamura2020Anyonic,
	author = {Nakamura, J. and Liang, S. and Gardner, G. C. and Manfra, M. J.},
	title = {{Direct observation of anyonic braiding statistics}},
	journal = {Nat. Phys.},
	volume = {16},
	pages = {931--936},
	year = {2020},
	month = sep,
	issn = {1745-2481},
	publisher = {Nature Publishing Group},
	doi = {10.1038/s41567-020-1019-1}
}

@article{Ofek2010PNAS,
	author = {Ofek, Nissim and Bid, Aveek and Heiblum, Moty and Stern, Ady and Umansky, Vladimir and Mahalu, Diana},
	title = {{Role of interactions in an electronic Fabry{\textendash}Perot interferometer operating in the quantum Hall effect regime}},
	journal = {Proc. Natl. Acad. Sci. U.S.A.},
	volume = {107},
	number = {12},
	pages = {5276},
	year = {2010},
	month = mar,
	doi = {10.1073/pnas.0912624107}
}

@article{Halperin2011PRB,
	author = {Halperin, Bertrand I. and Stern, Ady and Neder, Izhar and Rosenow, Bernd},
	title = {{Theory of the Fabry-P{\ifmmode\backslash\else\textbackslash\fi}'erot quantum Hall interferometer}},
	journal = {Phys. Rev. B},
	volume = {83},
	number = {15},
	pages = {155440},
	year = {2011},
	month = apr,
	issn = {2469-9969},
	publisher = {American Physical Society},
	doi = {10.1103/PhysRevB.83.155440}
}

@article{Nakamura2022NatCom,
	author = {Nakamura, J. and Liang, S. and Gardner, G. C. and Manfra, M. J.},
	title = {{Impact of bulk-edge coupling on observation of anyonic braiding statistics in quantum Hall interferometers}},
	journal = {Nat. Commun.},
	volume = {13},
	number = {344},
	pages = {1--9},
	year = {2022},
	month = jan,
	issn = {2041-1723},
	publisher = {Nature Publishing Group},
	doi = {10.1038/s41467-022-27958-w}
}

@article{Chklovskii1992Edge,
	author = {Chklovskii, D. B. and Shklovskii, B. I. and Glazman, L. I.},
	title = {{Electrostatics of edge channels}},
	journal = {Phys. Rev. B},
	volume = {46},
	number = {7},
	pages = {4026--4034},
	year = {1992},
	month = aug,
	issn = {2469-9969},
	publisher = {American Physical Society},
	doi = {10.1103/PhysRevB.46.4026}
}

@article{Nakamura2019NatPhys,
	author = {Nakamura, J. and Fallahi, S. and Sahasrabudhe, H. and Rahman, R. and Liang, S. and Gardner, G. C. and Manfra, M. J.},
	title = {{Aharonov{\textendash}Bohm interference of fractional quantum Hall edge modes}},
	journal = {Nat. Phys.},
	volume = {15},
	pages = {563--569},
	year = {2019},
	month = jun,
	issn = {1745-2481},
	publisher = {Nature Publishing Group},
	doi = {10.1038/s41567-019-0441-8}
}

@article{Roosli2020PRB,
	author = {R{\ifmmode\ddot{o}\else\"{o}\fi}{\ifmmode\ddot{o}\else\"{o}\fi}sli, Marc P. and Brem, Lars and Kratochwil, Benedikt and Nicol{\ifmmode\acute{\imath}\else\'{\i}\fi}, Giorgio and Braem, Beat A. and Hennel, Szymon and M{\ifmmode\ddot{a}\else\"{a}\fi}rki, Peter and Berl, Matthias and Reichl, Christian and Wegscheider, Werner and Ensslin, Klaus and Ihn, Thomas and Rosenow, Bernd},
	title = {{Observation of quantum Hall interferometer phase jumps due to a change in the number of bulk quasiparticles}},
	journal = {Phys. Rev. B},
	volume = {101},
	number = {12},
	pages = {125302},
	year = {2020},
	month = mar,
	publisher = {American Physical Society},
	doi = {10.1103/PhysRevB.101.125302}
}

@article{Roosli2021SciAdv,
	author = {R{\ifmmode\ddot{o}\else\"{o}\fi}{\ifmmode\ddot{o}\else\"{o}\fi}sli, Marc P. and Hug, Michael and Nicol{\ifmmode\acute{\imath}\else\'{\i}\fi}, Giorgio and M{\ifmmode\ddot{a}\else\"{a}\fi}rki, Peter and Reichl, Christian and Rosenow, Bernd and Wegscheider, Werner and Ensslin, Klaus and Ihn, Thomas},
	title = {{Fractional Coulomb blockade for quasi-particle tunneling between edge channels}},
	journal = {Sci. Adv.},
	volume = {7},
	number = {19},
	year = {2021},
	month = may,
	issn = {2375-2548},
	publisher = {American Association for the Advancement of Science},
	doi = {10.1126/sciadv.abf5547}
}

@article{Choi2015NatCom,
	author = {Choi, H. K. and Sivan, I. and Rosenblatt, A. and Heiblum, M. and Umansky, V. and Mahalu, D.},
	title = {{Robust electron pairing in the integer quantum hall effect regime}},
	journal = {Nat. Commun.},
	volume = {6},
	number = {7435},
	pages = {1--7},
	year = {2015},
	month = jun,
	issn = {2041-1723},
	publisher = {Nature Publishing Group},
	doi = {10.1038/ncomms8435}
}

@article{Sivan2018PRB,
	author = {Sivan, I. and Bhattacharyya, R. and Choi, H. K. and Heiblum, M. and Feldman, D. E. and Mahalu, D. and Umansky, V.},
	title = {{Interaction-induced interference in the integer quantum Hall effect}},
	journal = {Phys. Rev. B},
	volume = {97},
	number = {12},
	pages = {125405},
	year = {2018},
	month = mar,
	publisher = {American Physical Society},
	doi = {10.1103/PhysRevB.97.125405}
}

@article{McClure2009PRL,
	author = {McClure, D. T. and Zhang, Yiming and Rosenow, B. and Levenson-Falk, E. M. and Marcus, C. M. and Pfeiffer, L. N. and West, K. W.},
	title = {{Edge-State Velocity and Coherence in a Quantum Hall Fabry-P{\ifmmode\backslash\else\textbackslash\fi}'erot Interferometer}},
	journal = {Phys. Rev. Lett.},
	volume = {103},
	number = {20},
	pages = {206806},
	year = {2009},
	month = nov,
	publisher = {American Physical Society},
	doi = {10.1103/PhysRevLett.103.206806}
}

@article{Sivan2016NatCom,
  title = {Observation of interaction-induced modulations of a quantum Hall liquid’s area},
  volume = {7},
  ISSN = {2041-1723},
  url = {http://dx.doi.org/10.1038/ncomms12184},
  DOI = {10.1038/ncomms12184},
  number = {1},
  journal = {Nature Communications},
  publisher = {Springer Science and Business Media LLC},
  author = {Sivan,  I. and Choi,  H. K. and Park,  Jinhong and Rosenblatt,  A. and Gefen,  Yuval and Mahalu,  D. and Umansky,  V.},
  year = {2016},
  month = jul
}

@article{Zhang2009_PRB,
  title = {Distinct signatures for Coulomb blockade and Aharonov-Bohm interference in electronic Fabry-Perot interferometers},
  author = {Zhang, Yiming and McClure, D. T. and Levenson-Falk, E. M. and Marcus, C. M. and Pfeiffer, L. N. and West, K. W.},
  journal = {Phys. Rev. B},
  volume = {79},
  issue = {24},
  pages = {241304(R)},
  numpages = {4},
  year = {2009},
  month = {Jun},
  publisher = {American Physical Society},
  doi = {10.1103/PhysRevB.79.241304},
  url = {https://link.aps.org/doi/10.1103/PhysRevB.79.241304}
}

@article{Baer2014QPC,
  title = {Interplay of fractional quantum Hall states and localization in quantum point contacts},
  volume = {89},
  ISSN = {1550-235X},
  url = {http://dx.doi.org/10.1103/PhysRevB.89.085424},
  DOI = {10.1103/physrevb.89.085424},
  number = {8},
  journal = {Physical Review B},
  publisher = {American Physical Society (APS)},
  author = {Baer,  S. and R\"{o}ssler,  C. and de Wiljes,  E. C. and Ardelt,  P.-L. and Ihn,  T. and Ensslin,  K. and Reichl,  C. and Wegscheider,  W.},
  year = {2014},
  month = Feb 
}

@phdthesis{RoosliThesis,
  doi = {10.3929/ETHZ-B-000511370},
  author = {{Roosli,  Marc Philippe}},
  school = {ETH},
  title = {Quasiparticle tunneling between edge channels in quantum dots in the integer and fractional quantum Hall regimes},
  publisher = {ETH Zurich},
  year = {2021}
}

@article{Rosenow2007PRL,
  title = {Influence of Interactions on Flux and Back-Gate Period of Quantum Hall Interferometers},
  author = {Rosenow, B. and Halperin, B. I.},
  journal = {Phys. Rev. Lett.},
  volume = {98},
  issue = {10},
  pages = {106801},
  numpages = {4},
  year = {2007},
  month = {Mar},
  publisher = {American Physical Society},
  doi = {10.1103/PhysRevLett.98.106801},
  url = {https://link.aps.org/doi/10.1103/PhysRevLett.98.106801}
}

@article{Rajesh2019PRLNeutral,
  title = {Melting of Interference in the Fractional Quantum Hall Effect: Appearance of Neutral Modes},
  author = {Bhattacharyya, Rajarshi and Banerjee, Mitali and Heiblum, Moty and Mahalu, Diana and Umansky, Vladimir},
  journal = {Phys. Rev. Lett.},
  volume = {122},
  issue = {24},
  pages = {246801},
  numpages = {5},
  year = {2019},
  month = {Jun},
  publisher = {American Physical Society},
  doi = {10.1103/PhysRevLett.122.246801},
  url = {https://link.aps.org/doi/10.1103/PhysRevLett.122.246801}
}

@article{Inoue2014NatComNeutral,
  title = {Proliferation of neutral modes in fractional quantum Hall states},
  volume = {5},
  ISSN = {2041-1723},
  url = {http://dx.doi.org/10.1038/ncomms5067},
  DOI = {10.1038/ncomms5067},
  number = {1},
  journal = {Nature Communications},
  publisher = {Springer Science and Business Media LLC},
  author = {Inoue,  Hiroyuki and Grivnin,  Anna and Ronen,  Yuval and Heiblum,  Moty and Umansky,  Vladimir and Mahalu,  Diana},
  year = {2014},
  month = jun
}

@article{Inoue2013FracIQHE,
	author = {Inoue, Hiroyuki and Grivnin, Anna and Ofek, Nissim and Neder, Izhar and Heiblum, Moty and Umansky, Vladimir and Mahalu, Diana},
	title = {{Fractional charges in emergent neutral modes at the integer quantum Hall effect}},
	journal = {arXiv},
	year = {2013},
	month = oct,
	eprint = {1310.0691},
	doi = {10.48550/arXiv.1310.0691}
}

@article{Kane1995Edgerec,
  title = {Impurity scattering and transport of fractional quantum Hall edge states},
  volume = {51},
  ISSN = {1095-3795},
  url = {http://dx.doi.org/10.1103/PhysRevB.51.13449},
  DOI = {10.1103/physrevb.51.13449},
  number = {19},
  journal = {Physical Review B},
  publisher = {American Physical Society (APS)},
  author = {Kane,  C. L. and Fisher,  Matthew P. A.},
  year = {1995},
  month = May,
  pages = {13449–13466}
}

@article{Bid2010Edgerec,
  title = {Observation of neutral modes in the fractional quantum Hall regime},
  volume = {466},
  ISSN = {1476-4687},
  url = {http://dx.doi.org/10.1038/nature09277},
  DOI = {10.1038/nature09277},
  number = {7306},
  journal = {Nature},
  publisher = {Springer Science and Business Media LLC},
  author = {Bid,  Aveek and Ofek,  N. and Inoue,  H. and Heiblum,  M. and Kane,  C. L. and Umansky,  V. and Mahalu,  D.},
  year = {2010},
  month = jul,
  pages = {585–590}
}

@ARTICLE{Sabo201Edgerec,
  title     = "Edge reconstruction in fractional quantum Hall states",
  author    = "Sabo, Ron and Gurman, Itamar and Rosenblatt, Amir and Lafont,
               Fabien and Banitt, Daniel and Park, Jinhong and Heiblum, Moty
               and Gefen, Yuval and Umansky, Vladimir and Mahalu, Diana",
  journal   = "Nat. Phys.",
  publisher = "Springer Science and Business Media LLC",
  volume    =  13,
  number    =  5,
  pages     = "491--496",
  month     =  may,
  year      =  2017,
  language  = "en"
}

@article{SimonRosenow2015PRL,
  title = {Enhanced Bulk-Edge Coulomb Coupling in Fractional Fabry-Perot Interferometers},
  author = {von Keyserlingk, C. W. and Simon, S. H. and Rosenow, Bernd},
  journal = {Phys. Rev. Lett.},
  volume = {115},
  issue = {12},
  pages = {126807},
  numpages = {5},
  year = {2015},
  month = {Sep},
  publisher = {American Physical Society},
  doi = {10.1103/PhysRevLett.115.126807},
  url = {https://link.aps.org/doi/10.1103/PhysRevLett.115.126807}
}

@article{GoldsteinGefen2016PRL,
  title = {Suppression of Interference in Quantum Hall Mach-Zehnder Geometry by Upstream Neutral Modes},
  author = {Goldstein, Moshe and Gefen, Yuval},
  journal = {Phys. Rev. Lett.},
  volume = {117},
  issue = {27},
  pages = {276804},
  numpages = {6},
  year = {2016},
  month = {Dec},
  publisher = {American Physical Society},
  doi = {10.1103/PhysRevLett.117.276804},
  url = {https://link.aps.org/doi/10.1103/PhysRevLett.117.276804}
}

@article{Stern2010PRB,
  title = {Interference, Coulomb blockade, and the identification of non-Abelian quantum Hall states},
  author = {Stern, Ady and Rosenow, Bernd and Ilan, Roni and Halperin, Bertrand I.},
  journal = {Phys. Rev. B},
  volume = {82},
  issue = {8},
  pages = {085321},
  numpages = {9},
  year = {2010},
  month = {Aug},
  publisher = {American Physical Society},
  doi = {10.1103/PhysRevB.82.085321},
  url = {https://link.aps.org/doi/10.1103/PhysRevB.82.085321}
}

@article{Halperin1982QHE,
  title = {Quantized Hall conductance, current-carrying edge states, and the existence of extended states in a two-dimensional disordered potential},
  author = {Halperin, B. I.},
  journal = {Phys. Rev. B},
  volume = {25},
  issue = {4},
  pages = {2185--2190},
  numpages = {0},
  year = {1982},
  month = {Feb},
  publisher = {American Physical Society},
  doi = {10.1103/PhysRevB.25.2185},
  url = {https://link.aps.org/doi/10.1103/PhysRevB.25.2185}
}

@article{Rosenblatt2017,
  title = {Transmission of heat modes across a potential barrier},
  volume = {8},
  ISSN = {2041-1723},
  url = {http://dx.doi.org/10.1038/s41467-017-02433-z},
  DOI = {10.1038/s41467-017-02433-z},
  number = {1},
  journal = {Nature Communications},
  publisher = {Springer Science and Business Media LLC},
  author = {Rosenblatt,  Amir and Lafont,  Fabien and Levkivskyi,  Ivan and Sabo,  Ron and Gurman,  Itamar and Banitt,  Daniel and Heiblum,  Moty and Umansky,  Vladimir},
  year = {2017},
  month = Dec 
}

@article{wang2013PRLEdgerec,
  title = {Edge Reconstruction in the $\ensuremath{\nu}\mathbf{=}2/3$ Fractional Quantum Hall State},
  author = {Wang, Jianhui and Meir, Yigal and Gefen, Yuval},
  journal = {Phys. Rev. Lett.},
  volume = {111},
  issue = {24},
  pages = {246803},
  numpages = {5},
  year = {2013},
  month = {Dec},
  publisher = {American Physical Society},
  doi = {10.1103/PhysRevLett.111.246803},
  url = {https://link.aps.org/doi/10.1103/PhysRevLett.111.246803}
}

@book{Ihn2009,
  title = {Semiconductor Nanostructures},
  ISBN = {9780199534425},
  url = {http://dx.doi.org/10.1093/acprof:oso/9780199534425.001.0001},
  DOI = {10.1093/acprof:oso/9780199534425.001.0001},
  publisher = {Oxford University Press},
  author = {Ihn,  Thomas},
  year = {2009},
  month = Nov 
}

@article{Hanson2007,
  title = {Spins in few-electron quantum dots},
  volume = {79},
  ISSN = {1539-0756},
  url = {http://dx.doi.org/10.1103/RevModPhys.79.1217},
  DOI = {10.1103/revmodphys.79.1217},
  number = {4},
  journal = {Reviews of Modern Physics},
  publisher = {American Physical Society (APS)},
  author = {Hanson,  R. and Kouwenhoven,  L. P. and Petta,  J. R. and Tarucha,  S. and Vandersypen,  L. M. K.},
  year = {2007},
  month = Oct,
  pages = {1217–1265}
}

@article{vanHouten2005,
	author = {van Houten, H. and Beenakker, C. W. J. and Staring, A. A. M.},
	title = {{Coulomb-Blockade Oscillations in Semiconductor Nanostructures}},
	journal = {arXiv},
	year = {2005},
	month = aug,
	eprint = {cond-mat/0508454},
	doi = {10.48550/arXiv.cond-mat/0508454}
}

@phdthesis{Duprezthesis,
	author = {Corentin Duprez},
	title = {{H}elical edge transport and {F}abry-{P}érot interferometry in graphene quantum {H}all effect --- hal.science},
	howpublished = {\url{https://hal.science/tel-03524702/}},
	year = {2021},
    school = {Universite Grenoble Alpes},
	note = {[Accessed 12-05-2026]},
}

\onecolumngrid
\newpage

\clearpage

\section*{Supplementary Section}

\subsection*{SAMPLE FABRICATION}
The devices are fabricated from a GaAs/AlGaAs heterostructure with a two-dimensional electron gas (2DEG) located 138 nm below the top surface. The 2DEG electron density $n_e = 1.05\times10^{11}$ cm$^{-2}$ and mobility $\mu = 4.9\times10^6$ cm$^2/Vs$ are measured at 4 K in the dark. The mesa was defined using photolithography, followed by the removal of the unwanted regions using BCl$_3$ + Ar reactive-ion etching. The ohmic contacts were first defined with electron beam lithography (EBL), followed by evaporation of a 5 nm Ni/120 nm Ge/240 nm Au/80 nm Ni/20 nm Au metallic stack, which was then alloyed at 440 C for 80 seconds in an RTP machine. For the cavity gates, the entire sample was first coated with 25 nm of HfO2 using ALD. Then the gates were written with EBL, followed by the deposition of a 5 nm Ti/20 nm Au bilayer. Then, unwanted HfO$_2$ was removed from all the alloyed ohmic bonding pads. Finally, a 25 nm Ti/480 nm Au thick bilayer was deposited to make contact between the gates and the bonding pads. 

\subsection*{MEASUREMENTS}

The device was measured in a BlueFors dry dilution refrigerator with a base temperature of $\sim$ 10 mK. From a fit to a zero-bias Coulomb resonance, we extract an electronic temperature of \(T_e = 180 \pm 20~\mathrm{mK}\). Electronic filters are installed on all transport lines for electron thermalization. Differential conductance measurements are performed using a Zurich Instruments lock-in amplifier with a \(0.5~\mathrm{V}\) AC excitation at \(107.777~\mathrm{Hz}\), attenuated by a homemade \(10^{-5}\) voltage divider. A Yokogawa GS200 sets the DC-voltage bias. The AC and DC signals are combined using a homemade voltage adder. The current is measured through a current-to-voltage amplifier (K-tip variable gain transimpedance amplifier). A Yokogawa GS200 is used to apply a DC voltage to each gate. The device is cooled from room temperature with all gates and contacts grounded. 

\subsection*{CONSTANT INTERACTION MODEL}
\label{sec:dottheory}

We analyze the data within the commonly used constant-interaction model, in which the quantum Hall cavity is treated as a single conducting node characterized by a total capacitance \(C_\Sigma\). Within this approximation, electron-electron interactions are incorporated through a single electrostatic energy scale, while microscopic rearrangements of charge inside the cavity are not treated explicitly. The electrostatic energy of a cavity containing \(N\) electrons is written as~\cite{Ihn2009,Hanson2007,vanHouten2005}
\begin{equation}
U(N)=\frac{\left[-Ne+Q_0\right]^2}{2C_\Sigma},
\label{eq:elenergy}
\end{equation}
where \(Q_0\) denotes the externally induced charge, including the effect of gate voltages and other static offsets. The total capacitance is given by the sum of all capacitances between the cavity and external conductors,
\begin{equation}
C_\Sigma = C_{\mathrm{S}} + C_{\mathrm{D}} + C_{\mathrm{PG}} + \sum_i C_i,
\end{equation}
where \(C_{\mathrm{S}}\) and \(C_{\mathrm{D}}\) are the capacitances to the source and drain, \(C_{\mathrm{PG}}\) is the plunger-gate capacitance and \(\sum_i C_i\) accounts for additional gate capacitances, which we assume to be negligible in our case because of the large size of the cavity, and also because very negative voltages must be applied to the barriers to achieve the strong-charging limit, thus minimizing any direct effect these gates might have to the dot. 

In this description, only capacitances between the cavity and objects external to it enter \(C_\Sigma\). Internal capacitances within the cavity, such as the capacitance between compressible bulk and interfering edge, are not included as separate terms as long as the cavity is treated as a single node, which is a reasonable argument since our dot is always in the strong-charging regime. Likewise, the self-capacitance of the cavity is not added independently, but is already contained in the effective total capacitance \(C_\Sigma\) through its coupling to the surrounding conductors and dielectric environment. When the dot is weakly coupled to the leads, the total capacitance can be interpreted as the self-capacitance, and this is how we calculate the area of the defined cavity.

The electrostatic energy of an island containing N electrons is

\begin{equation}
    E_{\mathrm{el}}(N) = \frac{e^2N^2}{2C_\Sigma}+\varepsilon_N,
\end{equation}

where \(\varepsilon_N\) denotes the single-particle contribution. The corresponding addition energy to add one electron to the dot containing N electrons is therefore
\begin{equation}
E_{\mathrm{add}}(N)= E_{\mathrm{el}}(N+1) -  E_{\mathrm{el}}(N)
= \frac{e^2}{C_\Sigma} + \Delta \varepsilon_N,
\end{equation}
with \(\Delta \varepsilon_N=\varepsilon_{N+1}-\varepsilon_N\). In the limit where the single-particle level spacing is negligible compared to the interaction scale (which is true for the large, micron sized cavity we are studying), the addition energy is dominated by the charging term,
\begin{equation}
E_C \equiv \frac{e^2}{C_\Sigma}.
\end{equation}

We note that the term charging energy is often defined inconsistently in the literature. In the following, we refer to the addition energy as the charging energy \(E_C\), defined as the energy cost to add one electron to the cavity. In the constant-interaction model, the electrostatic energy of an island containing N electrons is given by Eq. \ref{eq:elenergy}, which contains a factor of \(1/2\). However, the addition energy corresponds to the discrete change \(E(N+1) - E(N)\), yielding \(E_C = e^2/C_\Sigma\) without the \(1/2\) factor (which is what we extract from the Coulomb diamond apex)~\cite{Ihn2009,Hanson2007}.

Generally, any of the capacitance terms entering \(C_\Sigma\) may depend on magnetic field,
\begin{equation}
C_\Sigma(B)=C_{\mathrm{S}}(B)+C_{\mathrm{D}}(B)+C_{\mathrm{PG}}(B),
\end{equation}
so that the charging energy becomes
\begin{equation}
E_C(B)=\frac{e^2}{C_\Sigma(B)}.
\end{equation}
Within this framework, a magnetic-field dependence of the measured charging energy reflects a magnetic-field dependence of the effective electrostatic environment seen by the cavity.

We emphasize that the capacitance extracted from the charging energy should be regarded as an \emph{effective} capacitance within the constant-interaction description. In the weak-coupling limit, $C_\Sigma$ can be interpreted as the electrostatic capacitance of the cavity to its environment. However, as the coupling to the leads increases, the measured addition energy may also be influenced by charge fluctuations and other interaction effects associated with the cavity--lead coupling. Consequently, the capacitance inferred from $E_C=e^2/C_\Sigma$ should be viewed as an effective parameter that captures the energy cost of adding an electron to the cavity. Throughout this work, when referring to magnetic-field-dependent changes in capacitance, we therefore mean changes in this effective capacitance as extracted from the measured charging energy.

\clearpage

\subsection*{CHARGING ENERGY EXTRACTION}
\label{sec:chenergy}

\begin{figure*}[htb!]
 \renewcommand{\thefigure}{S1}
 \includegraphics[width = 0.9\textwidth]{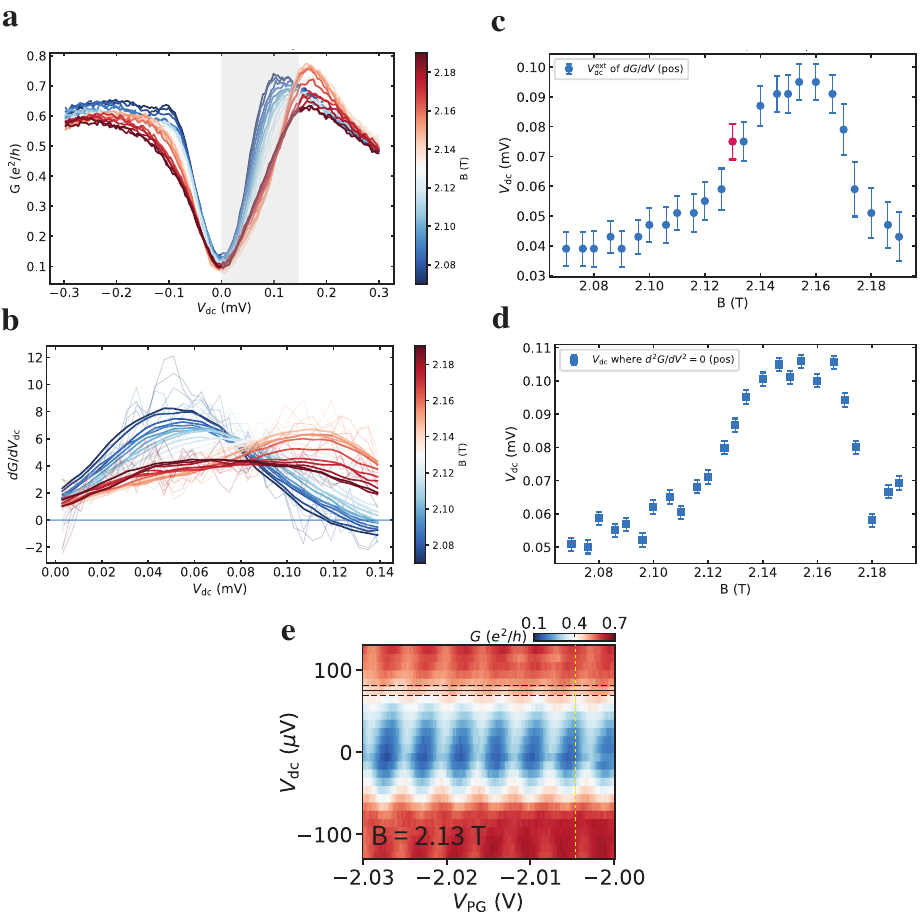}
 \begin{center}
 \caption{\textbf{Extraction of $V_{DC}(B)$ for Fig.~\ref{fig:2}h.} \textbf{a}, Linecuts along the DC bias direction for different magnetic fields.  The grey window indicates the $V_{DC}$ window where the derivative of \textbf{b} is displayed. \textbf{b}, First derivative of the linecuts in \textbf{a} with respect to $V_{DC}$. Raw derivative traces are shown as faint semi-transparent lines, while moving mean smoothed derivatives are shown as bold solid curves. \textbf{c}, Extracted $V_{DC}(B)$ obtained from the position of the maximum in the first derivative for each magnetic field in \textbf{b}. \textbf{d}, Extracted $V_{DC}(B)$ obtained from the zero crossing of the second derivative, providing an independent verification of the result in \textbf{c}. \textbf{e}, Coulomb diamonds. The black solid line marks the extracted $V_{DC}$ at B=2.13 T, obtained from the first-derivative analysis. This is given by the red point in \textbf{c}. The yellow dashed line indicates the plunger-gate voltage along which the DC-bias linecuts were taken. Black dashed lines denote the uncertainty bounds on the extracted $V_{DC}(B)$.}
 \label{sup:1}
 \end{center}
\end{figure*}

To determine the extent of bias of the Coulomb diamond (charging energy) as a function of magnetic field from $V_{\mathrm{DC}}-B$ measurements, we analyze linecuts of the differential conductance $G(V_{\mathrm{DC}})$ at fixed $B$ values (see Fig.~\ref{sup:1}a). For each linecut, the diamond apex (boundary) is defined as the bias position corresponding to the steepest change in conductance.
 
This $V_{DC}$ value is extracted from the extremum of the first derivative $dG/dV_{\mathrm{DC}}$, taking the maximum of the first derivative on the positive-bias side of $V_{DC}$. We focus exclusively on the positive DC-bias branch, since, as shown in Fig.~\ref{fig:2}b,f, a vertical line cut in DC bias at fixed plunger-gate voltage intersects only the positive apex of the Coulomb diamond. Due to the asymmetric source--drain bias application, the diamonds acquire a finite slope, such that the negative-bias branch does not probe the corresponding charging transition in an equivalent manner. It is therefore meaningful to restrict the analysis to the positive branch.

Following differentiation of the data, a light moving mean filter is applied to accurately select the extrema of the derivative (see \ref{sup:1}b). This is repeated for all linecuts along the magnetic field direction, and the resulting $V_{\mathrm{DC}}(B)$ is shown in Fig.~\ref{sup:1}c. The selected extrema are cross-checked against the nearest zero crossing of the second derivative $d^2G/dV_{\mathrm{DC}}^2$, shown in Figure \ref{sup:1}d, which marks the inflection point of the original trace and provides a consistency check that the identified feature corresponds to a genuine transition. 

This procedure implicitly assumes that the diamond edge manifests as a sharp step in $G(V_{\mathrm{DC}})$, such that the point of maximum slope accurately reflects the diamond boundary. In the presence of additional structure, such as excited states or overlapping resonances, this criterion may become ambiguous and lead to systematic deviations in the extracted bias values. However, only the first extremum of the first derivative (closest to $V_{DC} = 0$) is extracted, to make sure that we are indeed extracting the transition from the diamond and not a transition to an excited state, which would only appear at more positive $V_{DC}$. The analysis illustrated here is performed for the effective transmission of \(\sim 28\%\) discussed in the main text. Figure~\ref{sup:1}e reproduces the Coulomb diamonds shown in Fig.~\ref{fig:2}f. The solid black line denotes the extracted \(V_{\mathrm{DC}}\) at B = 2.13 T, identified in red in Fig.~\ref{sup:1}c, while the dashed lines indicate the corresponding uncertainty bounds.

The uncertainty in the extracted bias is estimated from the width of the derivative peak near its maximum. Specifically, we determine the voltage range over which the derivative exceeds 90\% of its maximum value and use this width as an estimate of the uncertainty.

For the determination of the charging energy ($E_c (B) = eV_{DC}(B)$), we use the bias positions obtained from the extrema of the first derivative $dG/dV_{\mathrm{DC}}$. This choice is motivated by the fact that the first-derivative extremum is a more direct measure of the steepest conductance variation, while the second derivative is used as a consistency check. 

Finally, we make the explicit assumption that the Coulomb diamonds do not shift with the magnetic field, which is essential since our analysis assumes that the measurements always probe the apex of the same Coulomb diamond. We consider this assumption to be justified because, in the strongly Coulomb-dominated regime, the conductance oscillations exhibit no continuous evolution with magnetic field, as evidenced by the nearly vertical features in the \(V_{\mathrm{PG}}-B\) plane. In the Supplementary Information section ``Coulomb diamonds at various magnetic fields for device B'', we further verify this assumption by repeating Coulomb diamond measurements across the investigated magnetic-field range and confirming the absence of any appreciable shift of the diamonds.

\clearpage

\subsection*{ADDITIONAL DATA FOR DEVICE B}
\label{supp:deviceB}

\begin{figure*}[htb!]
 \renewcommand{\thefigure}{S2}
 \includegraphics[width = 0.9\textwidth]{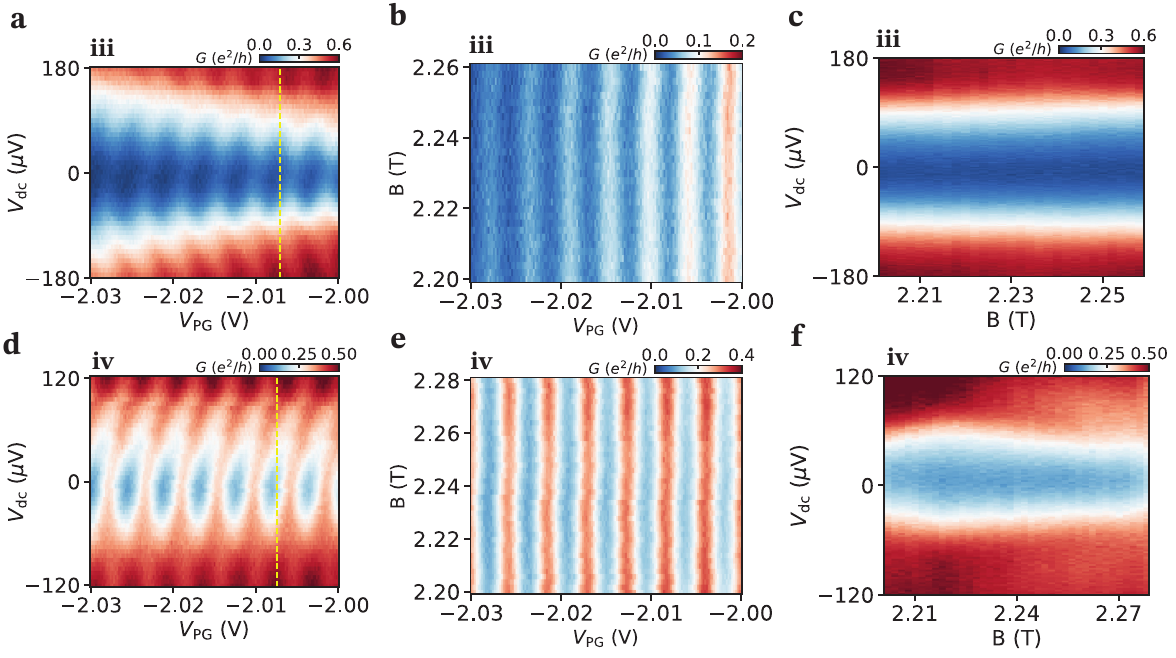}
 \begin{center}
 \caption{\textbf{a--c}, Data for configuration \textbf{iii} in the map given in Fig. \ref{sup:2b}. 
\textbf{d--f}, Data for configuration \textbf{iv} in the map given in Fig. \ref{sup:2b}. 
 \textbf{a,d}, Coulomb diamonds taken at 2.205 T; the yellow dashed line marks the fixed plunger-gate voltage used in \textbf{c,f}.  
\textbf{b,e}, Evolution of conductance oscillations with magnetic field.  
\textbf{c,f}, Conductance at fixed plunger-gate voltage (yellow dashed line in \textbf{a,d}), showing the evolution of a chosen diamond with magnetic field and DC bias. The DC bias is the fast axis in this measurement. }
 \label{sup:2a}
 \end{center}
\end{figure*}

\begin{figure*}[htb!]
 \renewcommand{\thefigure}{S3}
 \includegraphics[width = 0.9\textwidth]{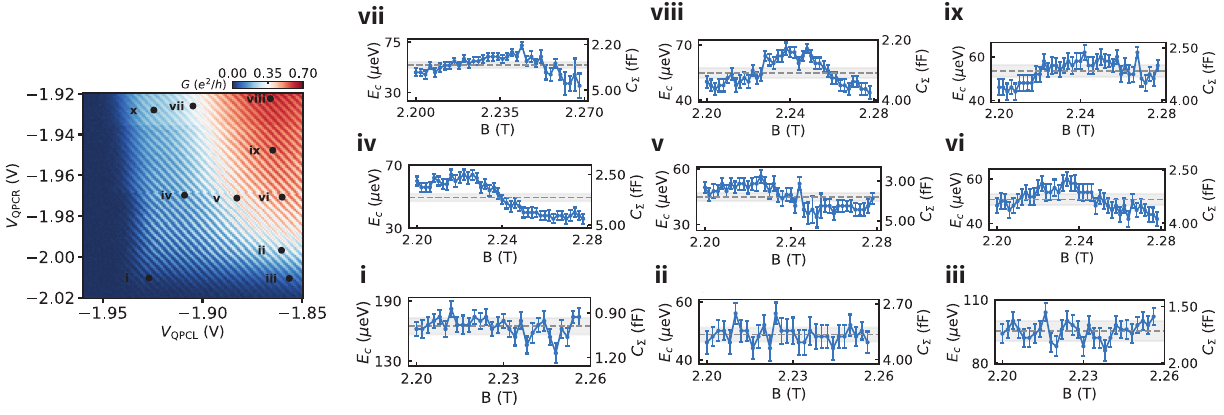}
 \begin{center}
 \caption{Extracted charging energy $E_{c}(B) = eV_{DC}(B)$ for all barrier combinations indicated in the left versus right QPC map, shown on the left.}
 \label{sup:2b}
 \end{center}
\end{figure*}

\clearpage

\subsection*{COULOMB DIAMONDS AT VARIOUS MAGNETIC FIELDS FOR DEVICE B}
\label{sec:diamond_shift}

\begin{figure*}[htb!]
 \renewcommand{\thefigure}{S4}
 \includegraphics[width = 0.9\textwidth]{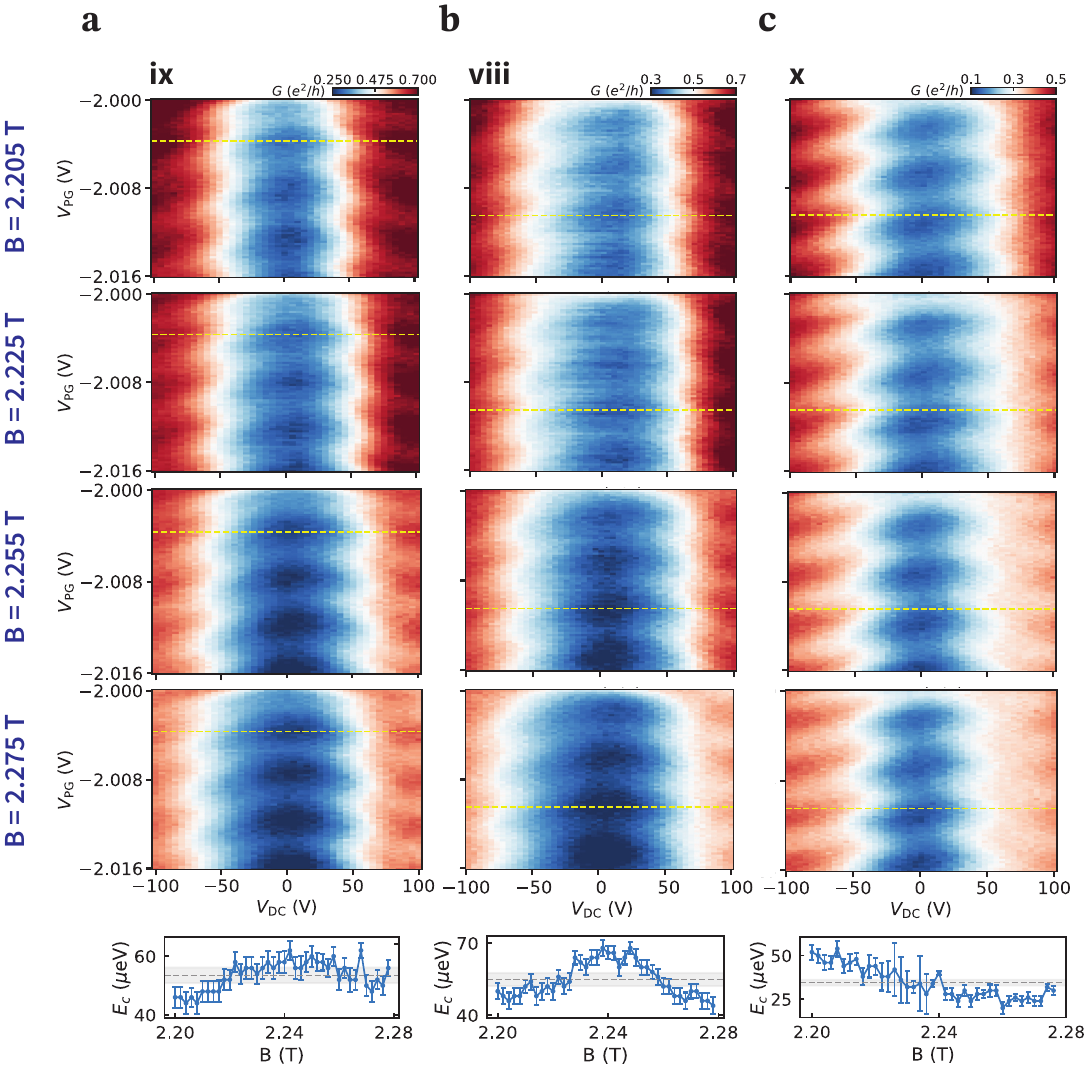}
 \begin{center}
 \caption{\textbf{Evolution of Coulomb diamonds with magnetic field.} Each column corresponds to the QPC configuration indicated by the corresponding Roman numeral in the map of Fig.~\ref{sup:2b}. The left column (\textbf{a}) corresponds to point \textbf{ix}, the middle column (\textbf{b}) to point \textbf{viii}, and the right column (\textbf{c}) to point \textbf{x}. The first four panels in each column show Coulomb diamonds measured at magnetic field values \(2.205\), \(2.225\), \(2.255\), and \(2.275~\mathrm{T}\), respectively. The yellow dashed line indicates the fixed plunger-gate voltage used to extract \(V_{\mathrm{DC}}(B)\). The final panel in each column shows the extracted \(E_c(B)\). Data sets such as those shown in \textbf{c} are excluded from the analysis because the diamonds tilt with the magnetic field, such that the fixed plunger-gate point no longer probes the diamond apex for all magnetic fields. We note that this is not a continuous shift of the coulomb oscillations, since for all QPC combinations we observe vertical lines in the $V_{\mathrm{PG}}-B$ plane (indicating that the diamonds do not shift appreciably with magnetic field), thus it is a tilt of the diamond apex as a function of magnetic field. We attribute this behavior to gate hysteresis occurring in specific QPC configurations. In contrast, for the configurations shown in \textbf{a} and \textbf{b}, we are probing how the apex, or charging energy, of a chosen diamond evolves with the magnetic field. All diamonds shown here were acquired with \(V_{\mathrm{DC}}\) as the fast scan axis, which was found to have fewer random charging events compared to sweeping the plunger-gate.}
 \label{sup:3}
 \end{center}
\end{figure*}

\subsection*{TEMPERATURE DEPENDENCE IN DEVICE B}
\label{sec:temp}

\begin{figure*}[htb!]
 \renewcommand{\thefigure}{S5}
 \includegraphics[width = 0.9\textwidth]{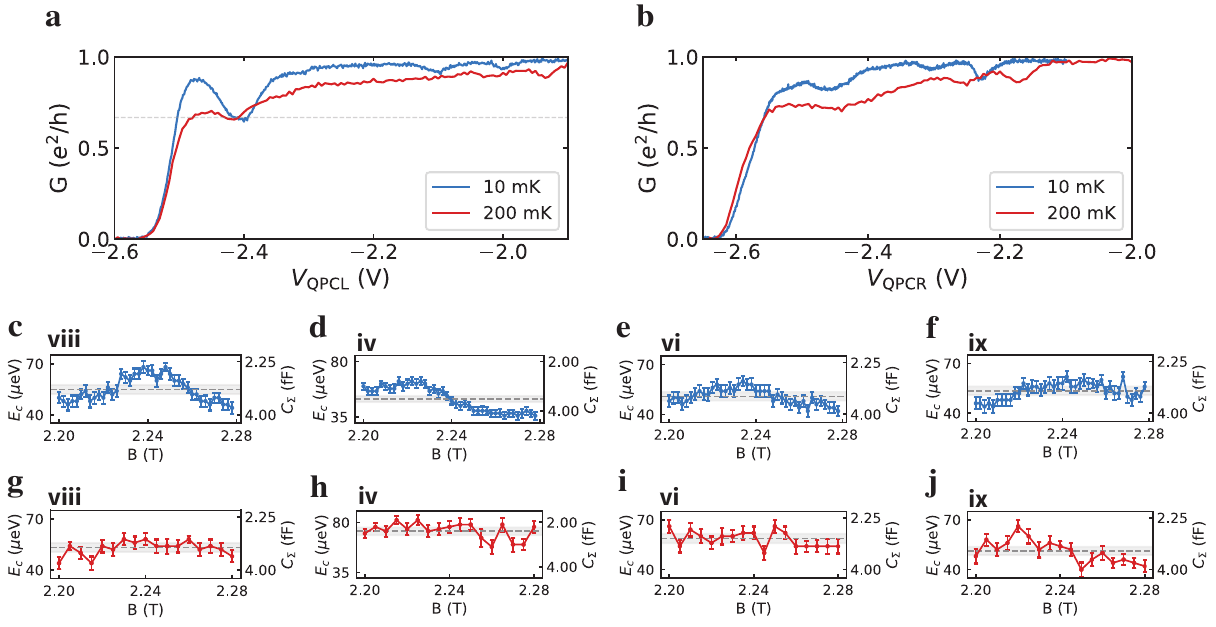}
 \begin{center}
\caption{\textbf{a}, Conductance of the left QPC measured at \(10~\mathrm{mK}\) (blue) and \(200~\mathrm{mK}\) (red). The grey dashed line indicates \(\nu = 2/3\). The plateau weakens at higher temperatures but remains visible. \textbf{b}, Conductance of the right QPC measured at \(10~\mathrm{mK}\) (blue) and \(200~\mathrm{mK}\) (red). In the following panels, the top row panel (blue) corresponds to data taken at \(10~\mathrm{mK}\), while the bottom row panel (red) corresponds to data taken at \(200~\mathrm{mK}\). \textbf{c,g}, Charging energy for region \textbf{viii}. \textbf{d,h}, Charging energy for region \textbf{iv}. \textbf{e,i}, Charging energy for region \textbf{vi}. \textbf{f,j}, Charging energy for region \textbf{ix}.}
 \label{sup:4}
 \end{center}
\end{figure*}

\end{document}